\documentclass[aps,prl,twocolumn,notitlepage,superscriptaddress]{revtex4-1}
\usepackage[colorlinks=true, citecolor=magenta, linkcolor=blue, urlcolor=blue]{hyperref}
\usepackage{graphicx}
\usepackage{amsmath}
\usepackage{amssymb}
\usepackage{amsfonts}
\usepackage{hyperref}
\usepackage{mathtools}
\usepackage{xcolor}
\usepackage{tikz}
\usepackage{enumitem}
\usetikzlibrary{shapes.geometric, arrows}
\usetikzlibrary{decorations.markings}

\usepackage[T1]{fontenc}
\usepackage{textcomp, amssymb}
\usepackage[utf8]{inputenc}

\allowdisplaybreaks

\begin{document}
 
\title{Enhancing Spin Transfer Torque in Magnetic Tunnel Junction Devices: Exploring the Influence of Capping Layer Materials and Thickness on Device Characteristics}

\author{Tahereh Sadat Parvini}
\email{tahereh.parvini@uni-greifswald.de}
\affiliation{Institut für Physik, Universität Greifswald, Greifswald, Germany}
\author{Elvira Paz}
\affiliation{INL - International Iberian Nanotechnology Laboratory, Avenida Mestre José Veiga, s/n, 4715-330 Braga, Portugal}
\author{Tim B\"ohnert}
\email{tim.bohnert@inl.int}
\affiliation{INL - International Iberian Nanotechnology Laboratory, Avenida Mestre José Veiga, s/n, 4715-330 Braga, Portugal}
\author{Alejandro Schulman}
\affiliation{INL - International Iberian Nanotechnology Laboratory, Avenida Mestre José Veiga, s/n, 4715-330 Braga, Portugal}
\author{Luana Benetti}
\affiliation{INL - International Iberian Nanotechnology Laboratory, Avenida Mestre José Veiga, s/n, 4715-330 Braga, Portugal}
\author{Felix Oberbauer}
\affiliation{Institut für Physik, Universität Greifswald, Greifswald, Germany}
\author{Jakob Walowski}
\email{jakob.walowski@uni-greifswald.de}
\affiliation{Institut für Physik, Universität Greifswald, Greifswald, Germany}
\author{Farshad Moradi}
\affiliation{ICELab, Aarhus University, Denmark}
\author{Ricardo Ferreira}
\affiliation{INL - International Iberian Nanotechnology Laboratory, Avenida Mestre José Veiga, s/n, 4715-330 Braga, Portugal}
\author{Markus M\"unzenberg}
\affiliation{Institut für Physik, Universität Greifswald, Greifswald, Germany}
\date{\today}

\begin{abstract}

We have developed and optimized two categories of spin transfer torque magnetic tunnel junctions (STT-MTJs) that exhibit a high tunnel magnetoresistance (TMR) ratio, low critical current, high outputpower in the micro watt range, and auto-oscillation behavior. These characteristics demonstrate the potential of STT-MTJs for low-power, high-speed, and reliable spintronic applications, including magnetic memory, logic, and signal processing. The only distinguishing factor between the two categories, denoted as A-MTJs and B-MTJs, is the composition of their free layers, 2 CoFeB/0.21 Ta/6 CoFeSiB for A-MTJs and 2 CoFeB/0.21 Ta/7 NiFe for B-MTJs. Our study reveals that B-MTJs exhibit lower critical currents for auto-oscillation than A-MTJs. We found that both stacks have comparable saturation magnetization and anisotropy field, suggesting that the difference in auto-oscillation behavior is due to the higher damping of A-MTJs compared to B-MTJs. To verify this hypothesis, we employed the all-optical time-resolved magneto-optical Kerr effect (TRMOKE) technique, which confirmed that STT-MTJs with lower damping exhibited auto-oscillation at lower critical current values. Additionally, our study aimed to optimize the STT-MTJ performance by investigating the impact of the capping layer on the device's response to electronic and optical stimuli.
\end{abstract}

\maketitle

%\subsection{Introduction}
The progress in deposition and crystal growth technologies has sparked significant interest in multilayer stacks, establishing them as a promising platform for advancing the fields of optical \cite{pantazopoulos2019layered, yu2021magnetic, parvini2015giant} and spintronic \cite{endoh2018recent, tanwear2020spintronic, stiewe2022} devices. Magnetic tunnel junctions (MTJs), are multilayer structures consisting of two ferromagnetic layers separated by an insulating barrier, and exhibit multifaceted functionality and efficacy across a broad range of scientific domains, including data storage, sensing, and computing \cite{zhu2006magnetic, tarequzzaman2018broadband, yang2018magnetic, maciel2019magnetic, jha2022interface}. Spin-transfer torque-based magnetic tunnel junctions (STT-MTJs) represent the next generation of MTJs, employing the spin-transfer torque effect to control the magnetization of a ferromagnetic layer. Compared to conventional MTJs, STT-MTJs offer superior performance characteristics, including higher speed, lower power consumption, and greater scalability. With the emergence of STT-MTJs, the potential of these devices has been significantly expanded, opening up new possibilities for their use in a wide range of applications \cite{parkin2008magnetic, ikeda2007magnetic, parkin2008magnetic, ikeda2007magnetic, jin2021highly, liu2002magnetic, wang2022picosecond, peng1999magnetic, sun2000spin, jin2021highly, liu2002magnetic, fong2016spin, farkhani2020lao}. The critical current density ($J_c$) in STT-MTJs refers to the minimum current density required to generate sufficient spin-transfer torque to switch the magnetization of the free layer. It is a key parameter in determining the performance and reliability of STT-MTJs, as it directly affects the switching speed and energy efficiency of the device \cite{verma2014modeling, diao2007spin, chun2012scaling, koopmans2005unifying, huang2020materials}. Since $J_c$ is proportional to the magnetic damping constant of STT-MTJs, materials and stacks with a low damping are desired for high-speed and low-power spintronic devices. 

In this study we developed two types of in-plane magnetized STT-MTJs, which were optimized in terms of layer arrangement, thicknesses, and size of the MTJ pillar to achieve a high TMR ratio and exhibit Vortex-based spin-torque nano-oscillator (STNO)-like behavior with low power consumption, i.e., a low critical current. The magnetization dynamics of the optimized STT-MTJs were scrutinized using time-resolved magneto-optical Kerr effect (TRMOKE) technique to extract key parameters, such as the damping coefficient and precession frequency as a function of the applied DC magnetic field.

%\subsection{Analysis of STT-MTJ Stacks: Structural Characterization and Electronic Response}
\begin{figure*}[ht]
  \centering
  \includegraphics[width=0.96\textwidth]{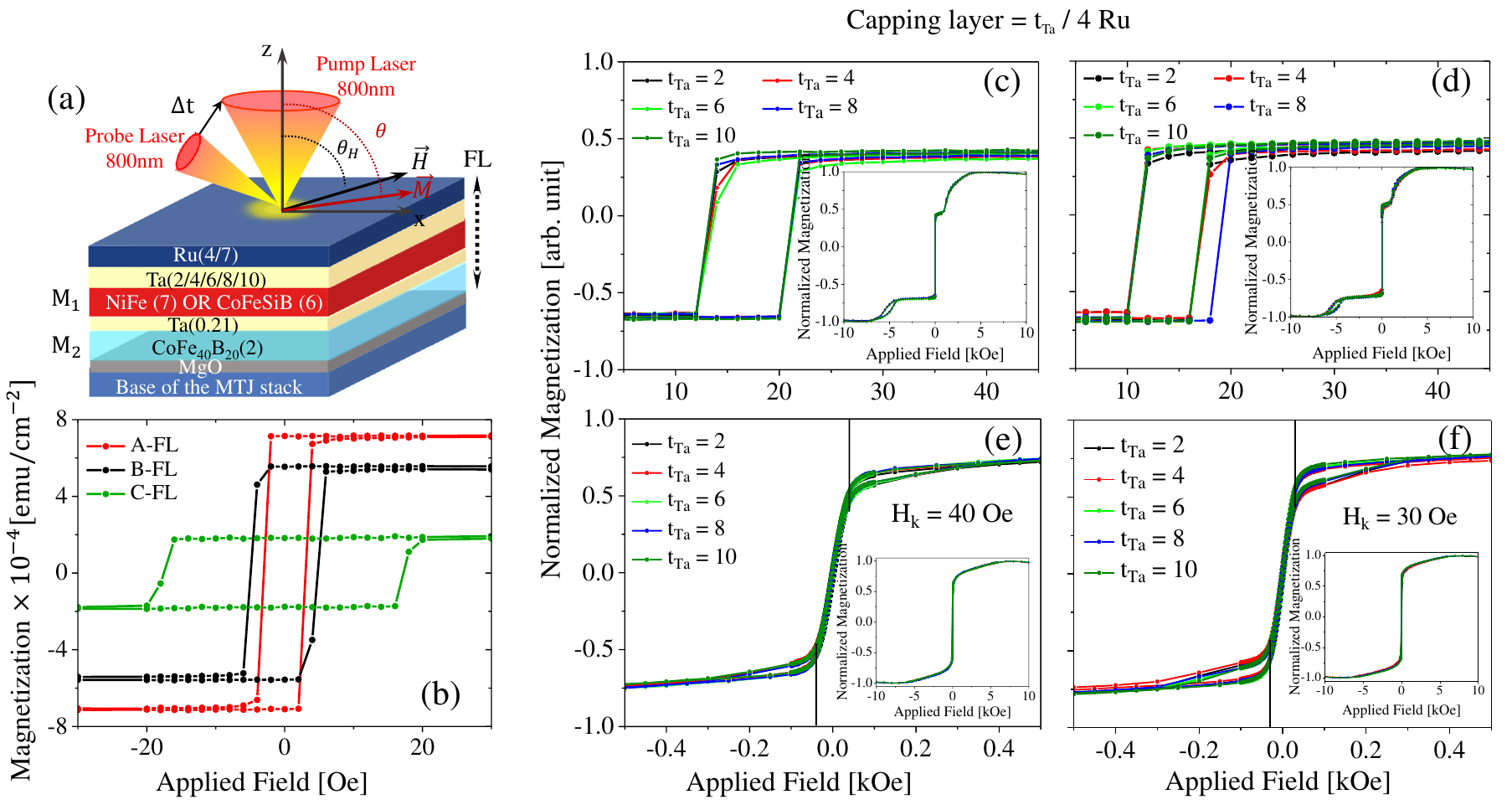}  
  
  \caption{(a) The illustration of the layer structure of optimized MTJs, with layer thicknesses indicated in nanometers, along with the geometry of ultrafast TRMOKE measurements. FL, $\Vec{H}$, and $\Vec{M}$ indicate the free layer of the MTJ, external applied magnetic field, and magnetization vector of the stack. (b) Hysteresis loops of the free layer stacks along the easy axis of magnetization. (c) and(d) Hysteresis loops along the easy-axis for full stacks B-MTJ and A-MTJ, respectively, and along the hard-axis for B-MTJ and A-MTJ in (e) and (f), respectively. The influence of varying thicknesses of the capping layer Ta $t_{\rm Ta}=$2/4/6/8/10 nm on the anisotropy field and hysteresis response has been studied. The value of $t_{\rm Ru}$ is consistently 4nm across all of the plots.}
\label{fig1}
\end{figure*}

A series of three distinct types of magnetic tunnel junctions (MTJs) were fabricated on SiO$_2$ (200nm) substrate using a singulus magnetron sputtering. As schematically depicted in Fig. \ref{fig1}(a), the MTJs were comprised of the following layers: 5 Ta / 50 CuN / 5 Ta / 50 CuN / 5 Ta / 5 Ru / 6 IrMn / 2 CoFe$_{30}$ / 0.825 Ru / 2.6 CoFe$_{40}$B$_{20}$ / MgO / Free Layer / Capping Layer, with thicknesses specified in nanometers. The free layer, which was the differentiating component among the MTJs, consists of 2 CoFe$_{40}$B$_{20}$ / X, where for type A-MTJ X$\equiv$0.21 Ta / 6 Co$_{67}$Fe$_{4}$Si$_{14.5}$B$_{14.5}$, for type B-MTJ X$\equiv$0.21 Ta / 7 Ni$_{80}$Fe$_{20}$ and for type C-MTJ X$\equiv$0. The thicknesses of CoFeSiB and NiFe in the A-MTJs and B-MTJs were meticulously chosen to guarantee comparable magnetic moments in both MTJs. Our strategy for selecting a trilayer configuration for the free layers in MTJs is as follows: Firstly, CoFeB was chosen to ensure a high tunneling magnetoresistance (TMR) ratio. Secondly, the inclusion of a Ta layer effectively prevented crystallinity issues between CoFeB and the NiFe (CoFeSiB) layer. Lastly, the incorporation of NiFe (CoFeSiB) as a soft ferromagnet facilitated the formation of a vortex within the free layer. Three alternatives were considered for the fabrication of the capping layer, namely Ta/Ru, Cu/Ru, and Ra/Ru. The elements Ta, Cu, and Ru are employed to inhibit the crystallization of the adjacent magnetic layer, while a few nanometers of Ru on top of the MTJ are utilized for the passivation of the stacks. For achieving maximum performance in MTJs, characterized by high TMR, low critical current, and sustained thermal and electrical stability, the optimization of both the material and thickness of the capping layer are crucial. To achieve this goal, we fabricated a series of A-MTJ and B-MTJ stacks with distinct capping layers and varied thicknesses. The analysis of Fig. \ref{fig1} and Fig. 1 SM \cite{SM}, measured by a MicroSense EV9 vibrating sample magnetometer (VSM), demonstrates that the magnetic response of MTJs remains mostly unaffected by the modification of the capping layer material. Furthermore, The hysteresis response of the stacks remains unaffected by the variation in thickness of Ta adjacent to the top magnetic layer in capping layers. Table \ref{magnetization} displays the saturation magnetization values ($M_\mathrm{s}$) of the distinct materials incorporated in the stack structures, namely CoFeSiB, CoFeB, NiFe, and CoFeSiB, as determined by VSM measurements. These values are needed to accurately analyze the magnetization dynamics of the MTJs in the last section of our work.

\begin{table}[h!]
    \caption{Saturation magnetization $M_\mathrm{s}$, in $10^3$A/m, of the materials: CoFe, CoFeB, NiFe, and CoFeSiB, measured by VSM.}
    \centering
    \begin{tabular}{l c c c c c c c c}
    \hline \hline
    Layer &&  CoFe$_{30}$ &&  CoFe$_{40}$B$_{20}$ &&  Ni$_{80}$Fe$_{20}$ &&  Co$_{67}$Fe$_{4}$Si$_{14.5}$B$_{14.5}$ \\ [1ex]
    \hline
    $M_\mathrm{s}$ && 1111.04 && 1358.98 && 680.61 && 1071.8 \\
    \hline \hline
\label{magnetization}
\end{tabular}
\end{table}

Furthermore, we synthesized simplified free layer stacks consisting of 5 Ta / 5 Ru / MgO / 2 CoFe$_{40}$B$_{20}$ / X / 2 Ta / 4 Ru, where X is defined as 
\begin{itemize}
  \item 0.21 Ta / 6 CoFe$_{40}$SiB$_{20}$ for A-FL, 
  \item 0.21 Ta / 7 Ni$_{80}$Fe$_{20}$ for B-FL,  
  \item and n/a for C-FL. 
\end{itemize}
These fabricated FL stacks facilitated a comparative investigation of their magnetization dynamics responses with their corresponding full stacks. As depicted in Figure \ref{fig1}(b), incorporating a low anisotropy NiFe layer results in nearly a 50$\%$ decrease in the coercivity of the free layer, which can be further reduced by replacing NiFe with an amorphous CoFeSiB layer.

\begin{figure}
    \centering
    \includegraphics[width=0.4\textwidth]{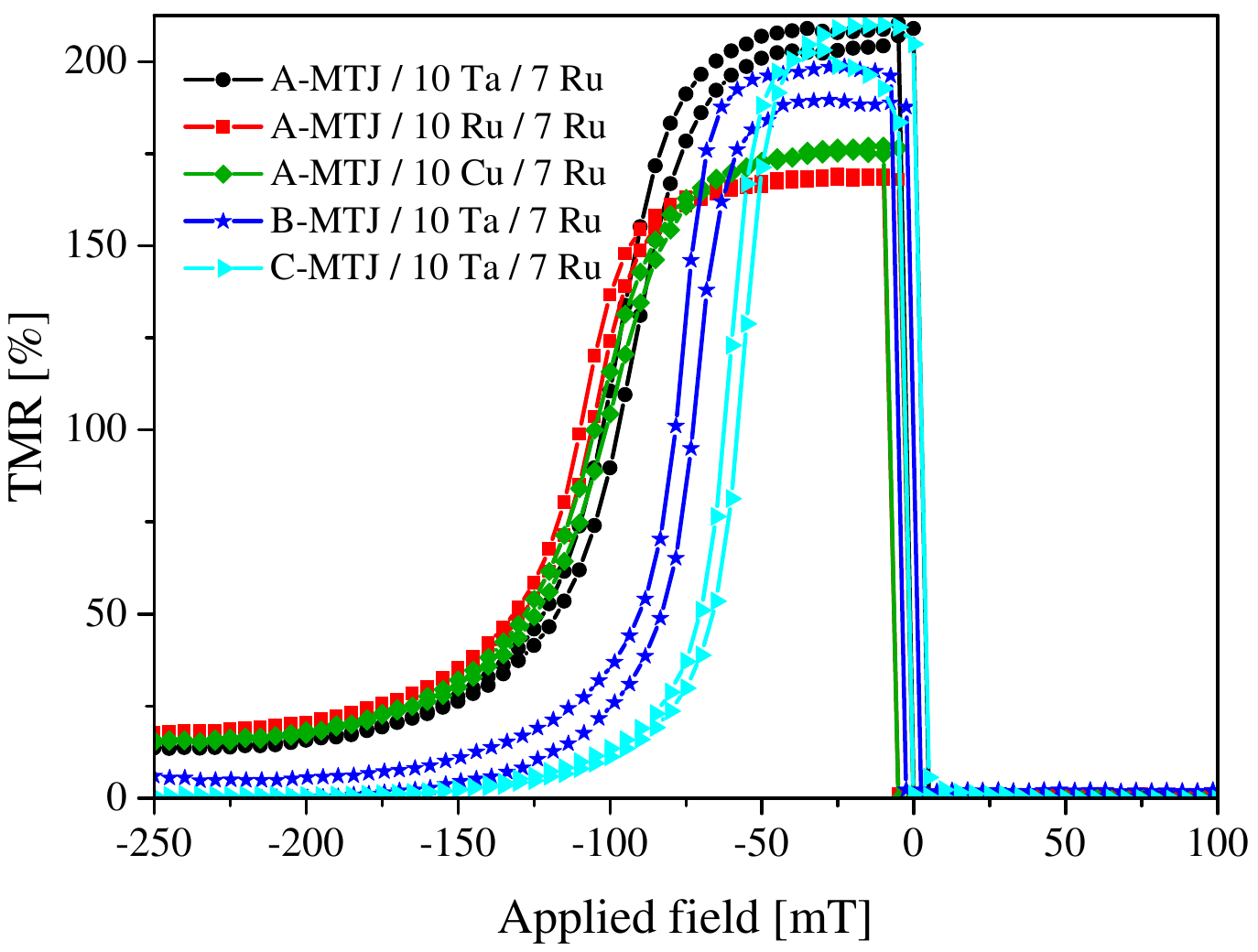}
    \caption{(a) TMR ratios of A-MTJ, B-MTJ, and C-MTJ with capping layers of various heavy metals measured using CIPT as a function of the applied magnetic field.}
\label{TMR}
\end{figure}

\begin{figure}
    \centering
    \includegraphics[width=0.47\textwidth]{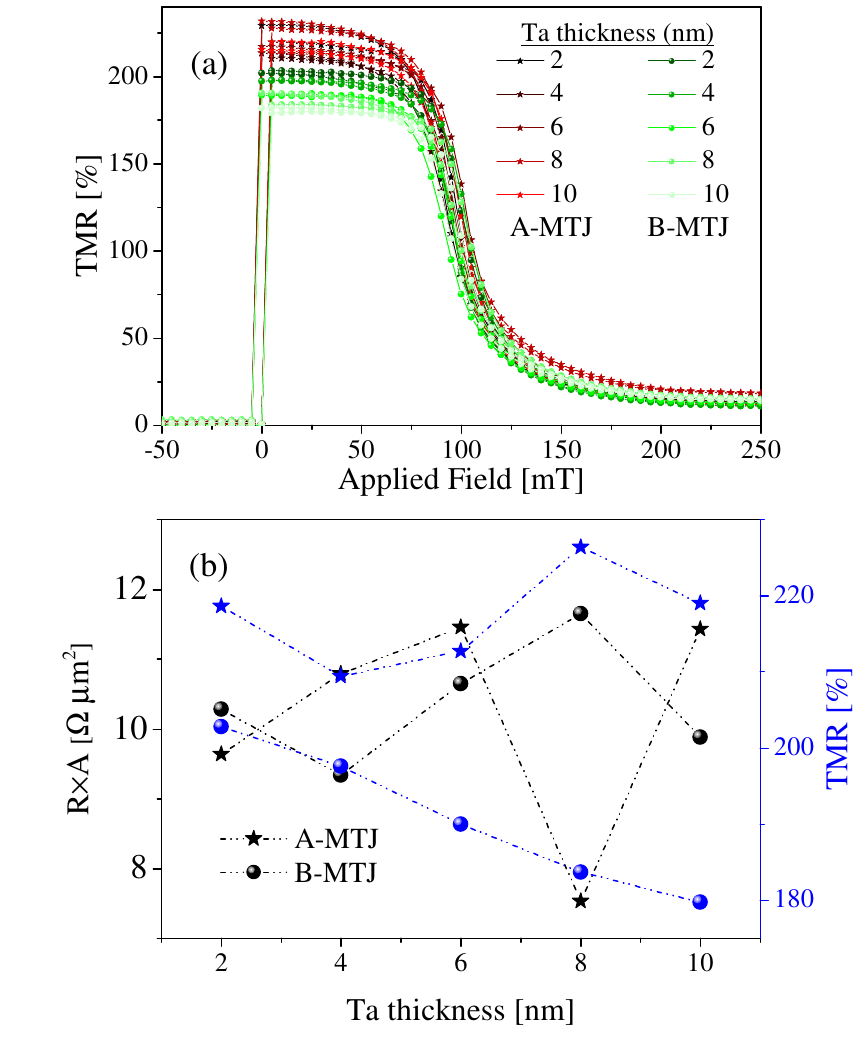}
    \caption{(a) TMR Ratio of MTJs versus Ta capping layer with capping layer X Ta/ 7 Ru. (b) Variation of TMR values and resistance area with Ta thickness. The dashed-dotted lines are guides to the eyes.}
\label{TMR2}
\end{figure}

The application of a direct electrical current in MTJs generates spin-transfer torques (STTs) that facilitate the transfer of angular momentum from the stationary polarizing magnetic layer to the free magnetic layer. This transfer results in oscillatory magnetization dynamics, which in turn produce an oscillatory electrical response via the magnetoresistance effect. The difference in resistance between parallel ($R_\mathrm{AP}$) and antiparallel ($R_\mathrm{P}$) magnetization configurations can be normalized to obtain the tunnel magnetoresistance ratio (TMR), expressed as TMR$=\frac{R_\mathrm{AP}-R_\mathrm{P}}{R_\mathrm{P}}$. The electrical properties of STT-MTJs were characterized using Current-In-Plane Tunneling (CIPT) measurements, and the resulting transfer curves for various stacks are presented in Figure \ref{TMR}. The TMR ratio of the B-MTJ stack is lower compared to the A-MTJs and resembles that of the control stack, C-MTJ. Our findings prove that the use of Ta in the capping layer of an MTJ stack leads to a higher TMR ratio compared to stacks utilizing Ru or Cu capping layers. We attribute this observation to the improved interface quality in stacks with Ta in the capping layer  which enhances spin-dependent transport. In contrast, Ru and Cu are prone to oxidation, which can lead to the formation of unwanted oxide layers and negatively impact the TMR performance \cite{nagahama2005spin, 1464430}. The exact measured values of the TMR ratio and resistance area ($R\times A$) of the aforementioned stacks are reported in Table \ref{CIPT}. In these STT-MTJs, the thickness of the MgO tunnel barrier is precisely selected to obtain a resistance area of approximately 10 $\mathrm{\Omega (\mu m)^2}$, which has been experimentally established as the optimal range for attaining maximum output power.

\begin{table}[h!]
    \caption{The TMR ratio and resistance-area values obtained from CIPT measurements for three different types of STT-MTJs.}
  \centering
    \begin{tabular}{l c c c c}
    \hline \hline
    MTJ Stacks &&&  TMR &  $R\times A$  \\
         &&&  ($\%$) &  $(\Omega \mu \mathrm{m}^2)$ \\[1ex]
    \hline
    B-MTJ / 10 Ta / 7 Ru &&& 190 & 11.2  \\
    A-MTJ / 10 Ta / 7 Ru &&& 208 & 9.6  \\
    A-MTJ / 10 Ru / 7 Ru &&& 168 & 10.3  \\
    A-MTJ / 10 Cu / 7 Ru &&& 179 & 11.0 \\
    C-MTJ / 10 Ta / 7 Ru &&& 199 & 11.9  \\
    \hline \hline
    \label{CIPT}
\end{tabular}
\end{table}

Having proven the superior performance of MTJs with the Ta capping layer, we embarked on an effort to determine the optimal thickness of the Ta layer for maximizing STT-MTJs' performance. As a consequence, we fabricate a collection of A-MTJs and B-MTJs, each with distinct thicknesses of Ta, namely $t_{\rm {Ta}}$=2/4/6/8/10 nm and a constant thickness of $t_{\rm Ru}$=7nm. The TMR ratio and resistance area of these stacks are shown in Fig. \ref{TMR2} (a) and (b). The TMR ratio of the B-MTJs exhibited a decrease as the thickness of Ta increased, whereas no noteworthy change in the TMR ratio of the A-MTJs was observed. Two potential reasons for the observed effect of Ta capping layer thickness on the TMR ratio are its role as a Boron sink, which is known to affect the TMR ratio through Boron diffusion, and its potential influence on the crystallization of the free layer. The crystallization of Ta depends on its thickness, with thin layers (<5 nm) crystallizing in beta-phase and thicker layers (>10 nm) in alpha-phase, which could affect the overall crystallization of the stack. However, the 0.21 nm Ta layer in the free layer is specifically designed to mitigate the influence of NiFe crystallization on CoFeB, which ultimately determines the TMR ratio.

\begin{figure*}[ht]
  \centering
  \includegraphics[width=0.8\textwidth]{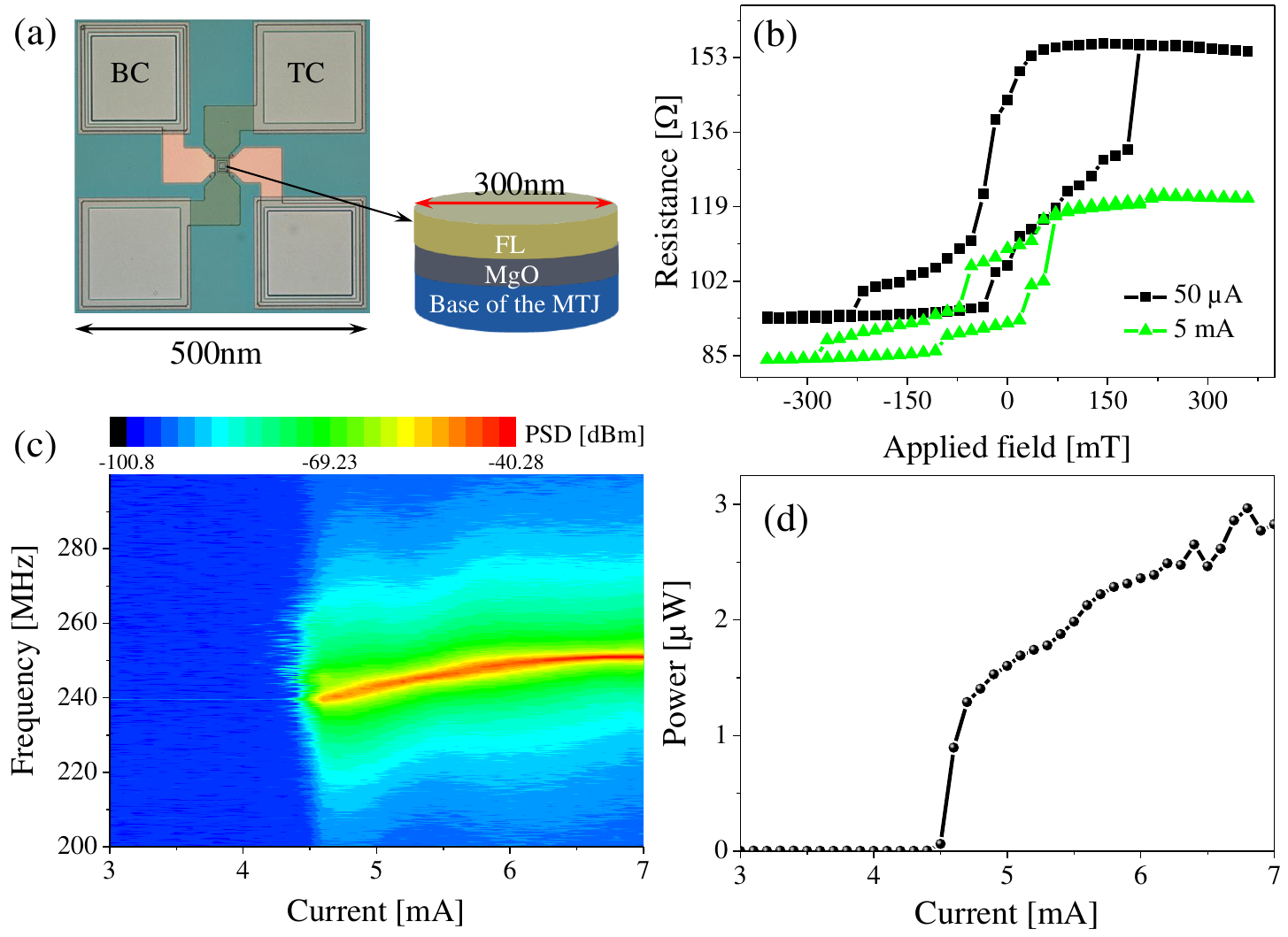}   
  \caption{(a) Micrograph of a device showing the B-MTJ with capping layer 10 Ta/7 Ru contacted via top and bottom contacts (TC and BC), along with a schematic representation of an B-MTJ nanopillar with a diameter of 300nm. (b) Resistance as a function of applied magnetic field for two applied currents, TMR= 66$\%$. (c) Dependence of the excited vortex frequency on the applied current. J$_c$=4.5 mA. (d) Dependence of the integrated output emitted power on the applied current.}
\label{fig4}
\end{figure*}

The minimum current density required to trigger the STT oscillations in the MTJs is critical current density, $J_{c}$.As shown in Fig. \ref{fig4}, for B-MTJ with a diameter of 300 nm critical current is about 4.5mA, or on the other hand around 0.064 $\mathrm{A}/\mu \mathrm{m}^2$, whereas no auto-oscillation was observed at low critical currents in the A-MTJs. Furthermore, Fig. \ref{fig4}(d) demonstrates that B-MTJ devices showcase a notable high output power in the micro-watt range. To theoretically validate these observations, we utilized the macrospin model, which provides an expression for critical current as \cite{khalsa2015critical, costa2017high, faber2009dynamic}
\begin{equation}
    J_{c}=\frac{2e\mu_{0}M_\mathrm{s}\alpha d}{\hbar}\left[\frac{M_\mathrm{s}}{2}+H+H_{k}\right].\frac{1}{P},
    \label{Jc}
\end{equation}
where $e$ is the charge of the electron, $\mu_0$ is the permeability of free space, $\hbar$ the Planck constant, $M_\mathrm{s}$ the magnetization saturation, $\alpha$ the Gilbert damping constant, $d$ the thickness of the free layer, and P the spin polarization. As previously described, both the B-MTJs and A-MTJs possess comparable saturation magnetization and anisotropy; thus, the relatively lower critical current of the former as compared to the latter can be attributed to their reduced damping. To validate this hypothesis, we utilized an TRMOKE technique method to measure the damping of the stacks. The MTJs with Ru and Cu capping layers not only demonstrated inferior performance when compared to those with a Ta capping layer, but also produced a noisier Kerr signal in TRMOKE measurements due to their elevated laser absorption and subsequently decreased penetration depth (under the same power of laser and magnetic field). Therefore in the following, only MTJs with capping layer X Ta/4 Ru are reported, due to their higher signal to noise ratio in the TRMOKE measurements. 
%\subsection{Investigating the Magnetization Dynamics of the optimized STT-MTJs}

\begin{figure}[h]
    \centering
   \includegraphics[width=.485\textwidth]{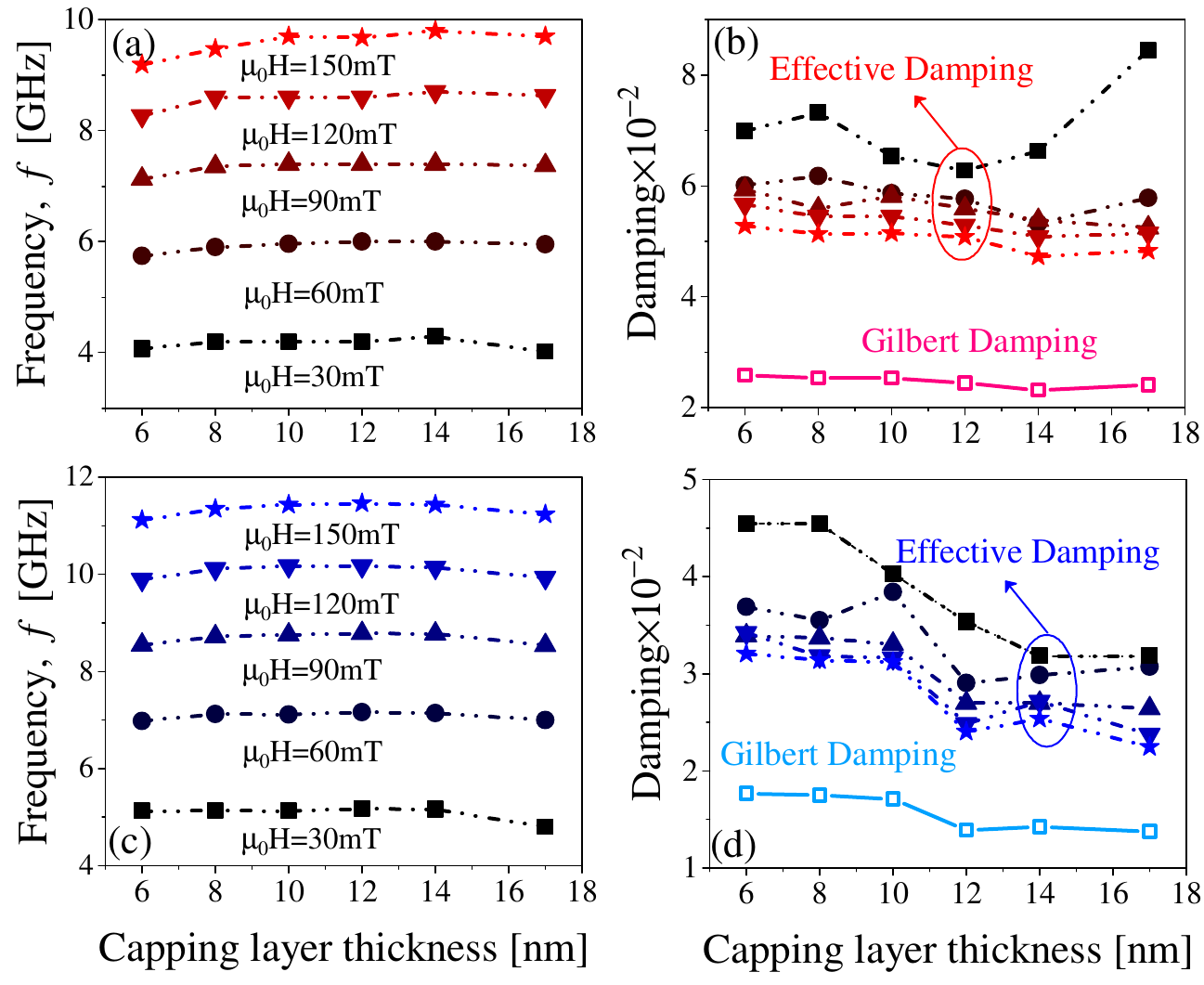}
  \caption{The precession frequency, effective damping constant ($\alpha_{eff}$), and Gilbert damping constant ($\alpha_{0}$) constants as a function of the capping layer thickness ($t_{\rm Ta}+t_{\rm Ru}$) for A-MTJs ((a) and (b)) and B-MTJs ((c) and (d)). The power of the pump is set to a constant value of 300 mW, while magnetic fields of 30, 60, 90, 120, and 150 mT (25$^{\circ}$ out of plane) are applied.}
\label{f_alpha}
\end{figure}

We employed an all-optical approach to investigate magnetization dynamics of the optimized STT-MTJs using femtosecond (fs) laser pulses in a pump-probe configuration (see Fig. \ref{fig1} (a)). The pump and probe pulses had central wavelengths of $\sim$800 nm, pulse durations of 40 fs, and repetition rates of 250 kHz. The pump and probe spot radii were 110 $\mu$m and 30 $\mu$m, respectively. Ultrafast demagnetization was induced by pump laser illumination, resulting in magnetization recovery and the damped precession of magnetization observed after a delay time of a few hundred femtoseconds. The static polar Kerr rotation angle $\theta_k$ was measured using the probe laser pulse. All TRMOKE measurements were performed at room temperature under an external magnetic field of up to 0.15 T at a fixed angle of $\theta_H=$75\textdegree{} measured from the film normal. The Supplementary Materials (SM) \cite{SM} contains documentation on the magnetization precession of free layer stacks and optimized full STT-MTJ stacks with varying capping layer thicknesses. The Fourier spectrum analysis of Kerr's data, as reported in \cite{SM}, reveals a single dominant precession frequency in all STT-MTJ, attributed to the direct ferromagnetic coupling of M$_1$ and M$_2$ through ultra-thin Ta layer and consequently their collective precession. Consequently, the macrospin model was utilized to derive the fitting function, enabling the determination of effective damping and precession frequency, and to solve the Landau–Lifshitz–Gilbert (LLG) equation, facilitating the determination of intrinsic damping in the MTJs. Then, TMOKE signals underwent a fitting process utilizing a damped-harmonic function with an exponential decay background and a sinusoidal term to achieve a meticulous analysis of the magnetization dynamics in stacks \cite{choi2020optical, choi2014spin, gonccalves2016dual}(detail in SM). Utilizing the applied fitting model, the precession frequencies ($f$) and magnetization relaxation times ($\tau$) corresponding to each STT-MTJS's macrospin were extracted. Fig. \ref{f_alpha} (a) and (c) reveals that the precession frequency of the macrospin in each stack is positively correlated with the magnetic field strength, with no noticeable impact from the thickness of the capping layer. Furthermore, the frequency of the full stack is lower than that of free layers, attributed to interlayer couplings. This interaction induces a shift in magnetic anisotropy and promotes a more stable configuration of magnetic moments, resulting in a reduction of the precession frequency. The effective damping coefficient of the macrospin vector denoted as $\alpha_{eff}$ and calculated using the formula $\alpha_{eff}=(2\pi f\tau)^{-1}$, is a measure of the total damping, encompassing both intrinsic Gilbert damping ($\alpha_{0}$) and extrinsic damping mechanisms. Under high fields, it suppresses extrinsic contributions and approaches the intrinsic Gilbert damping parameter, sees Fig. \ref{f_alpha} (b) and (d). The complexity of magnetization behavior and the greater contribution of inhomogeneous broadening make data collected at low magnetic fields particularly challenging to interpret. Therefore, to obtain a clearer understanding of the magnetization dynamics, high-field data is preferred. Overall, this figure indicates that B-MTJs exhibit both higher frequency and lower damping when compared to A-MTJs. The higher damping coefficient of A-MTJs compared to B-MTJs can be attributed to their amorphous structure and the presence of Silicon and Boron, which provide numerous energy-dissipating sites.

To extract the intrinsic Gilbert damping parameter using Gilbert's ansatz, it is essential to establish a connection between the exponential decay time $\tau$ and $\alpha_0$. Assuming negligible in-plane anisotropy and small tilting angles of the magnetization out of the sample plane ($\theta\approx\frac{\pi}{2}$), the precession frequency can be derived using Kittel's formula $\omega=\gamma\sqrt{\mu_{0}H_{x}\left[\mu_{0}H_{x}+\mu_{0}M_\mathrm{s}-\frac{2K_{z}}{M_\mathrm{s}}\right]}$ where $\omega$ is the angular frequency of the precession, $\gamma=\frac{g\mu_B}{\hbar}$ is the gyromagnetic ratio of MTJ stacks, $\mu_0$ and $\mu_B$ are the vacuum permeability and Bohr magneton respectively, g is the Landé g-factor, and $H_\mathrm{k}=2K_{z}/M_\mathrm{s}$ is the perpendicular anisotropy field \cite{khodadadi2017interlayer, zhang2022significant, wang2018temperature, walowski2008intrinsic}. By using the $M_\mathrm{s}$ and $H_\mathrm{k}$ data acquired from the VSM and utilizing a curve-fitting algorithm to match Kittel's formula with the corresponding frequencies derived from the TRMOKE, it is possible to determine the $\gamma$ (reported in Fig. 8(SM) \cite{SM}). Subsequently, the intrinsic Gilbert damping constant can be determined using the following equation:
\begin{equation}
    \alpha_0=\left[\frac{\tau\gamma}{2}\left(2\mu_{0}H_{x}+\mu_{0}M_\mathrm{s}-\frac{2K_{z}}{M_\mathrm{s}}\right)\right]^{-1}.
\end{equation}
The intrinsic Gilbert damping of stacks with different capping layers are shown in Fig. \ref{f_alpha}. This observation clearly illustrates the convergence of effective and intrinsic damping at high magnetic fields. Additionally, this observation confirms that B-MTJs exhibit lower intrinsic damping than A-MTJs, which correlates with the lower critical current density of B-MTJs. In accordance with TMR ratio measurements, our findings indicate that the damping of A-MTJ stacks displays negligible sensitivity to variations in capping layer thickness. Conversely, we observed a slight reduction in damping for B-MTJs as the capping layer thickness increased. This could be attributed to factors that contribute to the reduction in their TMR ratio, such as the potential impact of Ta on the crystallization of the free layer and its function as a Boron sink.
%\subsection{Conclusion}

By undertaking rigorous structural and dimensional optimization, we carefully designed two distinct categories of magnetic tunnel junctions, resulting in a combination of high TMR ratio, low critical current for auto-oscillation, and elevated output emitted power. All of these parameters are crucial for neuromorphic applications. The only point of distinction between the MTJ categories is the composition of their free layer, where category A-MTJs comprise 2 CoFeB / 0.21 Ta / 6 CoFeSiB, while category B-MTJs integrate 2 CoFeB / 0.21 Ta / 7 NiFe. Implementing CoFeSiB as the top magnetic layer led to a high TMR ratio of approximately $\simeq208\,$\%, whereas stacks containing NiFe exhibited a 15-20$\%$ lower TMR ratio. We observed that the critical current of B-MTJs was relatively lower than that of A-MTJs. Based on the nearly identical saturation magnetization and anisotropy field values measured by VSM, we inferred that the lower critical current of B-MTJs could be attributed to their lower Gilbert damping. We validated this hypothesis through TRMOKE measurements. The magnetic characteristics of MTJs with NiFe exhibit sensitivity to the thickness of the capping layer, likely due to the influence of Ta on the crystallization of the free layer and its role as a boron sink, whereas CoFeSiB-based MTJs do not show such behavior.\\
%\subsection{Acknowledgement}

This project has received funding from the European Union’s Horizon 2020 research and innovation program under grant agreement No 899559 (SpinAge).

\subsection{Availability of data}
The data that supports the findings of this study are available within the article and its supplementary material.

%\bibliography{main.bib}

\end{document}